\documentclass[aps, prb, twocolumn]{revtex4-2}

\usepackage{graphicx}
\graphicspath{{figures/}}
\usepackage[utf8]{inputenc}
\usepackage{amssymb, amsmath, commath, braket, physics}
\usepackage{hyperref}
\hypersetup{
colorlinks=true,
linkcolor=red,
filecolor=blue,
urlcolor=magenta,
}

\newcommand{\Ecal}{\ensuremath{\mathcal{E}}}	

\newcommand{\ferf}[1]{\ensuremath{\mathrm{Erf}\del{#1}}}
\newcommand{\ferfc}[1]{\ensuremath{\text{Erfc}\del{#1}}}

\newcommand{\fbslj}[2]{\ensuremath{\mathrm{J}_{#1}\del{#2}}}

\newcommand\gEM{ \ensuremath{\gamma_\mathrm{E}} }

\def\J{\ensuremath \mathsf{J}}					

\newcommand\lboro{School of science, Loughborough University, Loughborough, Leicestershire LE11 3TU, UK}

\newcommand\mean[1]{\ensuremath{\left\langle #1 \right\rangle}}

\newcommand{\nrm}{\ensuremath{Z}}

\def\ot{\ensuremath{\mathsf{O}}}				

\newcommand{\prob}[1]{\ensuremath{\textrm{P}\del{#1}}}

\def\rog{\ensuremath{\hat{\rho}_\mathrm{GE}}}		

\newcommand\tdip{\ensuremath{t_\mathrm{dip}}}
\newcommand\tH{\ensuremath{t_\mathrm{H}}}
\newcommand\tR{\ensuremath{t_\mathrm{R}}}
\newcommand\tZ{\ensuremath{t_\mathrm{Zeno}}}

\newcommand\Zsym{\ensuremath{ \mathbb{Z}_2 }}
\begin{document}

\title{Spectral statistics, non-equilibrium dynamics and thermalization in random matrices with global \Zsym-symmetry}
\author{Adway Kumar Das}
\affiliation{\lboro}

\begin{abstract}
	$\mathbb{Z}_2$ symmetry is ubiquitous in quantum mechanics where it drives various phase transitions and emergent physics. The role of $\mathbb{Z}_2$ symmetry in the thermalization of a local observable in a disordered system can be understood using random matrix theory. To do so, we consider random symmetric centrosymmetric (SC) matrix as a toy model where a $\mathbb{Z}_2$ symmetry, namely, the exchange symmetry is conserved. Such a conservation law splits the Hilbert space into decoupled subspaces such that the energy spectrum of a SC matrix is a superposition of two pure spectra. After discussing the known results on the correlations of such mixed spectrum, we consider different initial states and analytically compute the time evolution of their survival probability and associated timescales. We show that there exist certain low-energy initial states which do not decay over a very long timescales such that a measure zero fraction of random SC matrices exhibit spontaneous symmetry breaking. Later, we look at the equilibrium values of local observables like the density-density correlation, kinetic energy operator and compare them against the average values from the microcanonical and canonical ensembles. We find that when the observable violates (respects) the global symmetry of the Hamiltonian, the equilibrium value is independent (dependent) of the symmetry of the initial state. However, irrespective of such symmetry constraints, the fluctuations of the diagonal terms of the observables within microcanonical shells decay with system size such that the ansatz of eigenstate thermalization hypothesis remains valid. We show that the equilibrium value converges to the canonical average for all the observables and initial states, indicating that thermalization occurs despite the presence of a global symmetry.
\end{abstract}


\maketitle
\section{Introduction}
The emergence of statistical mechanics upon equilibration of an isolated quantum system is a fascinating area of research. Despite the unitarity of the time evolution of an isolated system, 
a typical local observable thermalizes if the energy states of the governing Hamiltonian are ergodic~\footnote{For a Hilbert space with dimension $N$, an ergodic state is uniformly distributed in the $N-1$-dim unit hypersphere~\cite{Neumann2010, Goldstein2010}.} with correlated energy spectrum~\cite{Srednicki1994, Polkovnikov2011a, Nandkishore2015, Borgonovi2016, Deutsch2018}. Surprisingly, many strongly correlated interacting many-body systems violate thermalization due to breaking of ergodicity~\cite{TorresHerrera2017a, Pino2016, Vidmar2016, Mori2018}, measure zero athermal initial states e.g.~quantum scars~\cite{Turner2018, Sinha2020, Dong2023, Pizzi2025}, kinetic constraints in glassy systems~\cite{Roy2020, Royen2024, Bhore2023, Das2026Arxiv}, emergent integrals of motion from many-body localization~\cite{Serbyn2013, Shtanko2025, Kulshreshtha2018}. 
Violation of thermalization is also observed in single-particle systems~\cite{Lydzba2021}, where approximately conserved charges can form due to on-site disorder or inhomogeous hopping terms~\cite{Modak2016, Das2023, Das2022, Das2024, Das2025b}. Thus, it is important to understand the role of conservation laws, hence, the underlying symmetries of a system in the statistical description of its equilibrium properties.

In a two-level system, the presence of a single non-trivial symmetry ensures integrability, where the energy levels become uncorrelated with localized eigenstates~\cite{Das2019, Das2023a, Berry2009}. This naturally leads to the question of what happens in a disordered system with many degrees of freedom upon breaking a single global symmetry~\cite{Das2022b, Das2022a}. In this work, we focus on the discrete $\mathbb{Z}_2$ symmetry where the Hamiltonian commutes with an invertible symmetry operator $\hat{\ot}$ possessing only two distinct non-zero eigenvalues. $\mathbb{Z}_2$ symmetry is ubiquitous in physics, e.g.~spontaneous breaking of a $\mathbb{Z}_2$ symmetry leads to quantum phase transition in transverse field Ising~\cite{Dutta2015book}, coupled top~\cite{Mondal2020, Wang2021}, Lipkin-Meshkov-Glick~\cite{Castanos2006, Santos2016}, Dicke~\cite{PerezFernandez2011, Kloc2018, Hwang2015} models. The degenerate spectrum of $\hat{\ot}$ implies that an $N$-dimensional Hilbert space with $N\gg 1$ can be split into two decoupled subspaces labeled by two different good quantum numbers. Consequently, any excitation initially localized on a particular subspace will always be restricted there if the governing Hamiltonian commutes with the symmetry operator $\hat{\ot}$.

If the symmetry imposed by $\hat{\ot}$ is slightly perturbed, the initially decoupled subspaces get weakly coupled. Then, starting from one subspace, a localized excitation can leak to the complimentary subspace with a finite probability. This leads to the thermalization of a generic local observable at sufficiently long time. But, within some intermediate time window, the non-equilibrium dynamics may exhibit long-lived metastable states, known as pre-thermalization~\cite{Kinoshita2006, Hofferberth2007, Gring2012, Langen2015, Mori2018, Mallayya2021}. 
This has important ramification in various problems, e.g.~breaking parity in nuclear physics \cite{Krupchitsky2014book}, isospin mixing~\cite{Mitchell1988, Guhr1990a, Adams1998}, 
exciton transfer~\cite{Scholak2011, Zech2014, Ortega2016}. 
Such explicit symmetry breaking not only affects the properties of quantum mechanical but of classical systems as well, e.g.~acoustic resonances in anisotropic quartz block~\cite{Ellegaard1996, Schaadt2003, Carvalho2007}, 
small-world~\cite{Carvalho2009}, biological~\cite{Jalan2009} networks. Other than breaking symmetry, weakly coupled subspaces can also form due to emergent phenomena, e.g.~quantum Zeno effect~\cite{Facchi2002}, cooperative shielding due to long-range interactions~\cite{Santos2016a}. 

As a concrete example of $\mathbb{Z}_2$ symmetry in a disordered system, we consider the random symmetric centrosymmetric (SC) matrices, which commute with the exchange symmetry operator, $\J$. Such a discrete symmetry is important in diverse disciplines ranging from information theory~\cite{Gersho1969, Magee1973} to engineering problems~\cite{Datta1989, Gaudreau2016}. In the construction of a pre-engineered quantum wire (i.e.~a network of qubits), exchange symmetry is shown to play a central role to ensure perfect transfer of information between the input and output qubits~\cite{Christandl2004, Karbach2005, Shi2005, CamposVenuti2007, Nikolopoulos2004, Faria2025}. The exchange symmetry is also important for the transfer fidelity in the photosynthetic structures~\cite{Scholak2011, Zech2014} and disordered networks~\cite{Ortega2016, Das2022b}.

In this work, we take random SC matrices and look at the dynamics of various initial states which are either confined to a particular symmetry sector or couples the symmetry sectors with a tunable parameter. In case of such initial states, we look at the equilibrium values of various local observables and compare them with the predictions from different statistical ensembles. Our system size scaling analyses indicate that thermalization occurs in case of a generic observable despite the presence of a global symmetry and the equilibrium value is described by the Gibbs ensemble~\cite{Rigol2007, Gogolin2011, Caux2013, Vidmar2016, Mallayya2021}.

The organization of the paper is as follows: in Sec.~\ref{sec:rsc}, we define the exchange symmetry, SC matrices and discuss some of their general properties. In Sec.~\ref{sec:SC_energy}, we provide the analytical expressions of the energy correlations in case of a random SC matrix. In Sec.~\ref{sec:SC_dynamics}, we show the analytical expression of the survival probability of certain initial states and discuss the characteristic timescales. Importantly, we show that spontaneous symmetry breaking can happen in a measure zero fraction of the random SC matrices. In Sec.~\ref{sec:SC_thermal}, we discuss the general notion of thermalization in an isolated system and the equilibration of various observables w.r.t.~different initial states in case of the SC matrices. Our concluding remarks are given in Sec.~\ref{sec:discuss}.

\section[Centrosymmetry]{Symmetric Centrosymmetric Matrix}\label{sec:rsc}

An $N\times N$ centrosymmetric matrix, $H$, satisfies the commutation relation 
\begin{align}
	[H, \J ] = 0 \Rightarrow H_{i,j} = H_{i', j'}\;\forall\; i,j,\quad i' \equiv N+1-i
\end{align}
where the exchange matrix $\J\equiv \cbr{\delta_{i, N+1-j}}$ is symmetric, invertible ($\J = \J^{-1} = \J^T$), involutory ($\J^2 = \mathbb{I}$, the identity matrix, has two distinct eigenvalues, $\pm 1$ with multiplicities $\lceil \frac{N}{2} \rceil$ and $\lfloor \frac{N}{2} \rfloor$, respectively. If $N = 2^L$, an $N\times N$ exchange matrix can be expressed as $\underset{k}{\otimes} \sigma^x_k$ where $\sigma^x_k$ is the Pauli X matrix in $\hat{\sigma}_z$ representation acting on the $k$th site of a fictitious 1D lattice with $L$ sites. Since $\sbr{\sigma^x, \mathbb{I}} = 0$, any matrix of the form $\sigma^x_i \sigma^x_j\dots \sigma^x_k$ commutes with $\J$ where $1\leq i<\dots < k \leq L$. The exchange matrix also commutes with the matrices of the form
\begin{align}
	\mathbb{I}^{N_1\times N_1} \oplus \ot\oplus\mathbb{I}^{N_2\times N_2}\oplus \J \ot\J \oplus \mathbb{I}^{N_1\times N_1}
\end{align}
where $\ot$ is any arbitrary operator of dimension $N - (2N_1+N_2)$.

A matrix $H$ with both centrosymmetry and diagonal symmetry is called symmetric centrosymmetric (SC)
\begin{align}
	H = \J H^T\J \Leftrightarrow H_{i,j} = H_{j', i'}
\end{align}
which is persymmetric as well~\cite{Cantoni1976}. Since $[H,\J ]=0$, we can reduce $H$ in the basis of $\J $ via a similarity transformation $Q^THQ = H'$ such that $H'$ assumes a block diagonal form. 
Such a block structure is a consequence of the two disjoint subspaces in the Hilbert space, 
which are denoted by odd and even sectors, spanned by anti-symmetric ($\ket{\Phi^-}$) and symmetric ($\ket{\Phi^+}$) states with energy levels $E^\mp$, respectively~\cite{Das2022b}.

\subsection{Energy correlations of a SC matrix}\label{sec:SC_energy}
\begin{figure}[t]
	\centering
	\includegraphics[width=0.95\columnwidth]{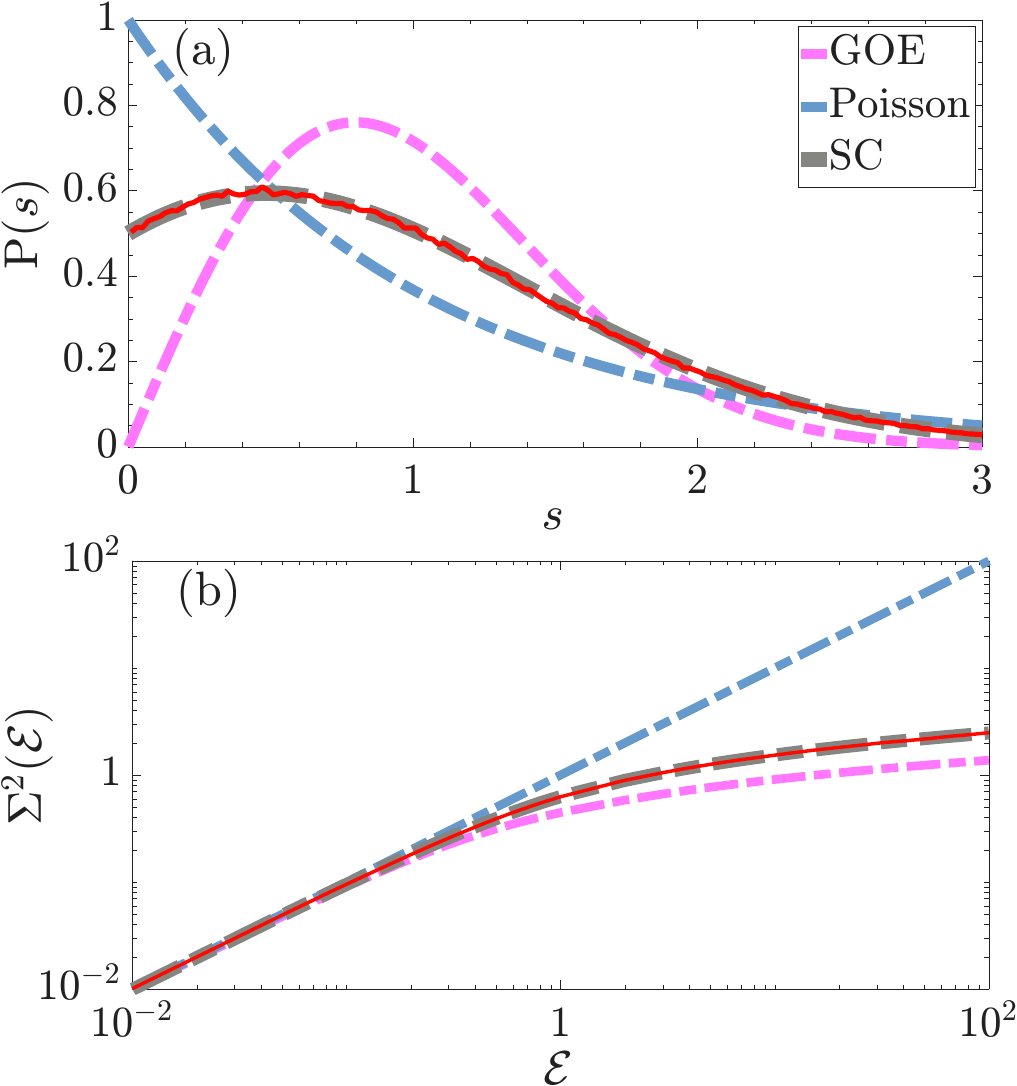}
	\caption[Energy correlation of SC matrices]{Energy correlation of random SC matrices. (a) density of level spacing (b) number variance for $N = 2048$. The solid curves denote numerical results while the dashed lines denote the analytical expressions for Poisson, GOE and random SC matrices (Eqs.~\eqref{eq_NNS_SC} and \eqref{eq_nvar_SC}).}
	\label{fig_SC_s_numvar}
\end{figure}
We look at random SC matrices having independent and identically distributed Gaussian random elements with zero mean and unit variance. Corresponding energy spectrum is a superposition of two pure spectra, each belonging to a Gaussian orthogonal ensemble (GOE) matrix. The energy correlations in case of such mixed spectrum is well known~\cite{Pandey2019Arxiv}. Below, we provide the analytical forms of the energy correlations for a random SC matrix for the sake of completeness.

The density of states of a random SC matrix follows the Wigner's semicircle law with a width $2\sqrt{\mean{E^2}} = \sqrt{4 + N}$. Corresponding energy levels exhibit level clustering with the density of level spacing~\cite{Mehta2004book, Pandey2019Arxiv}
\begin{align}
	\label{eq_NNS_SC}
	\mathrm{P}^\mathrm{SC}(s) = \frac{e^{-\frac{\pi}{8}s^2}}{2} + \frac{\pi}{8} s e^{-\frac{\pi}{16}s^2} \ferfc{\frac{\sqrt{\pi}}{4}s}
\end{align}
and the 2nd moment of the level spacing is
\begin{align}
	\mean{s^2} = \frac{8(2 - \sqrt{2})}{\pi}.
\end{align}
In Fig.~\ref{fig_SC_s_numvar}(a), we plot the numerically obtained density of level spacing of SC ensemble for $N = 2048$ along with the analytical expression from Eq.~\eqref{eq_NNS_SC}. The above equation indicates a degree of level repulsion intermediate between that of the GOE and Poisson ensemble ($\mean{s^2} = \frac{4}{\pi}$ and 2 for the Wigner's surmise and exponential distribution, respectively). An intermediate level repulsion caused by incomplete separation of symmetries is observed in many-body systems as well~\cite{Santos2010}.

The superposition of two pure spectra also affects the long-range energy correlation of a SC matrix, reflected in the two-point correlation function, 
\begin{align}
	\label{eq_def_rho2}
	\rho^{(2)}(E_1, E_2) = N(N-1) \int dE_3\dots dE_N \prob{\vec{E}}.
\end{align}
The above integral relation is valid provided the joint density of the $N$ energy levels, $\prob{\vec{E}}$ is invariant under the exchange of any pair of energies. The two-point correlation, 
$\rho^{(2)}(E_1, E_2)$ is the probability of finding any two energy levels at $E_1$ and $E_2$~\cite{Bogomolny2001}. 
A related quantity is the two-level cluster function, $T_2(E) \equiv 1 - \rho^{(2)}(E)$ where $E = E_1 - E_2$. 

To understand the universal features of the energy correlations, it is important to unfold the energy spectrum, $\Ecal_j = N \varrho(E_j)$ where $\varrho(x)$ is the cumulative distribution of energy~\cite{Guhr1998}. Unfolding gets rid of the system dependent global shape of the density of states and reduces the mean level spacing to unity. We denote the two-level cluster function of the unfolded energy levels as $Y_2(\Ecal)$ where $\Ecal$ is the gap of unfolded energy levels. In the Poisson ensemble, $Y_2(\Ecal) = 0$ at any energy gap, while for GOE~\cite{Mehta2004book}
\begin{align}
	\label{eq_Y2_GOE}
	\begin{split}
		Y_2^\mathrm{GOE}(\Ecal) &= \del{\frac{\pi}{2}\mathrm{sgn}(\Ecal) - \mathrm{Si}(\pi \Ecal)} \del{\frac{\cos\pi \Ecal}{\pi \Ecal} - \frac{\sin \pi \Ecal}{(\pi \Ecal)^2}}\\ &+ \del{\frac{\sin \pi \Ecal}{\pi \Ecal}}^2
	\end{split}
\end{align}
where $\mathrm{sgn}(x)$ is the Signum function and $\mathrm{Si}(x) \equiv \int_{0}^{x} dt\frac{\sin t}{t}$ is the sine integral. In case of a SC matrix, the superposition of two GOE spectra leads to a two-level cluster function~\cite{Pandey2019Arxiv}
\begin{align}
	\label{eq_Y2_SC}
	Y_2^\mathrm{SC}(\Ecal) = \frac{1}{2} Y_2^\mathrm{GOE}\del{\frac{\Ecal}{2}}.
\end{align}

The inverse Fourier transform of the two-level cluster function gives the two-level form factor, a function in the dimensionless time domain, $\tau$
\begin{align}
	\label{eq_def_b2}
	b_2(\tau) = \int_{-\infty}^{\infty} d\Ecal\: Y_2(\Ecal) \cos(2\pi \tau \Ecal)
\end{align}
where the symmetry of $Y_2(\Ecal)$ around $\Ecal = 0$ reduces the complex integral to a real one~\cite{Buijsman2020}. The two-level form factor controls the long time dynamics of a generic observable~\cite{Schiulaz2019}. For Poisson ensemble, $b_2(\tau) = 0$ at any time while for GOE
\begin{align}
	\label{eq_b2_GOE}
	b_2^{\mathrm{GOE}}(\tau) &= \begin{cases}
		1 - 2\tau + \tau\log\del{1 + 2\tau}, & \tau\leq 1,\\
		\tau\log \left( \dfrac{2\tau + 1}{2\tau - 1} \right) - 1, &\tau > 1 .
	\end{cases}
\end{align}
Then, Eq.~\eqref{eq_Y2_SC} implies that the two-level form factor of a SC matrix is $b_2^\mathrm{SC}(\tau) = b_2^{\mathrm{GOE}}(2\tau)$.

To identify different energy scales in a system, one of the most useful measure is the number variance, defined for a set of unfolded energies, $\cbr{\Ecal_j}$ as~\cite{Das2025a},
\begin{align}
	\begin{split}
		\Sigma^2(\Ecal) &= \mean{ [\mathcal{N}(\Ecal, \Ecal_0) - \Ecal]^2}_{\Ecal_0}\\
		\mathcal{N}(\Ecal, \Ecal_0) &= I\del{\Ecal_0 + \frac{\Ecal}{2}} - I\del{\Ecal_0 - \frac{\Ecal}{2}}
	\end{split}
	\label{eq_nvar}
\end{align}
where $I(\Ecal)$ is the cumulative distribution of the unfolded energy levels and $\mathcal{N}(\Ecal, \Ecal_0)$ is the number of energy levels in the window of span $\Ecal$ and centered at $\Ecal_0$. For GOE, the energy spectrum is rigid so that any two energy levels are correlated and the number variance has a logarithmic behavior~\cite{Mehta2004book}
\begin{align}
	\label{eq_num_var_GOE}
	\begin{split}
		\Sigma^2_\mathrm{GOE}(\Ecal) = \frac{2}{\pi^2} \log\del{ 2\pi \Ecal + \gEM + 1 - \frac{\pi^2}{8} } + \mathcal{O}(\frac{1}{\Ecal})
	\end{split}
\end{align}
where $\gEM = 0.577216\dots$ is the Euler-Mascheroni constant, $\mathrm{Ci}(x) = -\int_{x}^{\infty}dt \frac{\cos t}{t}$ and $\mathrm{Si}(x) = \int_{0}^{x}dt \frac{\sin t}{t}$. Contrarily in the Poisson ensemble, the number of unfolded energy levels are free to fluctuate around their mean position, $\mean{\mathcal{N}(\Ecal, \Ecal_0)}_{\Ecal_0} = \Ecal$ and the number variance exhibits a linear behavior, $\Sigma^2(\Ecal) = \Ecal$. In case of a SC matrix, superposition implies that the number variance follows~\cite{Pandey2019Arxiv}
\begin{align}
	\label{eq_nvar_SC}
	\Sigma^2_\mathrm{SC}(\Ecal) = 2 \Sigma^2_\mathrm{GOE}\del{\frac{\Ecal}{2}}
\end{align}
In Fig.~\ref{fig_SC_s_numvar}(b), we plot the numerically obtained number variance of SC ensemble for $N = 2048$ along with the analytical expression from Eq.~\eqref{eq_nvar_SC} and find an excellent agreement. Thus, we have closed form analytical expressions of both the short- and long-range energy correlations of the SC matrices.

\section{Dynamical response of a SC matrix}\label{sec:SC_dynamics}
The exchange symmetry of a SC matrix implies that the dynamics of an initial state will depend on its symmetry. To illustrate this, we look at the following initial state
\begin{align}
	\label{eq_ini_state_SC}
	\ket{\Psi_\omega} \equiv \cos \omega \ket{k} + \sin \omega \ket{k'}, \quad -\frac{\pi}{2} \leq \omega \leq \frac{\pi}{2}
\end{align}
where $k' \equiv N+1-k$ and $\ket{k}$ is any unit vector from the computational basis. The expectation value of the exchange symmetry w.r.t.~$\ket{\Psi_\omega}$ is
\begin{align}
	\mean{\J} = \bra{\Psi_\omega} \J \ket{\Psi_\omega} = \sin (2\omega).
	\label{eq:J_avg}
\end{align}
Our choice of the initial state is motivated by Schr\"odinger's cat states, which are superposition of macroscopically different states~\cite{Ourjoumtsev2006, Huang2006, Wang2016, Lewenstein2021}. In particular, $\ket{\Psi_\omega}$ is a weighted superposition of the exchange doublets, $\frac{\ket{k}\pm \ket{k'}}{\sqrt{2}} \equiv \ket{\Psi_{\pm\frac{\pi}{4}}}$, which are eigenstates of $\J$ and have energies $H_{k,k}\pm H_{k,k'}$ w.r.t.~the Hamiltonian $H$.

The local density of states~\cite{Tavora2016, Schiulaz2019, Jana2021, Louis2023, Prasad2024} of $\ket{\Psi_\omega}$ is
\begin{align}
	\begin{split}
		\rho_\omega(E) &= \sum_{j = \pm }\sum_{n = 1}^{\frac{N}{2}} |c_n^j|^2 \delta\del{E - E_n^j}\\
		c_n^\pm &\equiv \bra{\Phi_n^\pm}\ket{\Psi_\omega} = (\cos\omega \pm \sin\omega)\Phi_n^\pm(k) 
	\end{split}
	\label{eq:LDOS_def}
\end{align}
where $\ket{\Phi_n^\pm}$ are eigenstates of $H$ with energies $E_n^\pm$. In case of the SC matrices, $\rho_\omega(E)$ follows the Wigner's semicircle law with a width $\Gamma \approx \frac{\sqrt{N}}{2}$ irrespective of $\omega$. However, for $\omega = \frac{\pi}{4}$, $\ket{\Psi_\omega}$ is a symmetric state (having $\mean{\J} = 1$), hence, spanned by only $\ket{\Phi^+_j}$'s, the symmetric eigenstates of $H$. Consequently, the time evolution of $\ket{\Psi_\omega}$ is always confined to the even subspace. Similarly, the time evolution of $\ket{\Psi_{-\frac{\pi}{4}}}$ is always restricted to the odd subspace spanned by anti-symmetric energy eigenstates, $\ket{\Phi^-}$'s. Contrarily for $\omega = 0$, $\ket{\Psi_\omega}$ couples both the odd and even subspaces and homogeneously spreads over the entire Hilbert space at sufficiently long time.

To understand the non-equilibrium features of the initial state $\ket{\Psi_\omega}$, we monitor its survival probability~\cite{Polkovnikov2011a, Torres2018, Torres2019, Schiulaz2019, Das2025, VallejoFabila2024, VallejoFabila2025}, defined as
\begin{align}
	\label{eq_R_t_def}
	\begin{split}
		R(t) &\equiv \abs{\bra{\Psi_\omega(t)}\ket{\Psi_\omega}}^2 = \abs{ \sum_{j = \pm }\sum_{n = 1}^{\frac{N}{2}} |c_n^j|^2 e^{-i E_n^j t} }^2
	\end{split}
\end{align}
Then, ensemble averaged survival probability can be expressed as~\cite{Schiulaz2019}
\begin{align}
	\label{eq_R_t_avg}
	\begin{split}
		\mean{R(t)} &= \mean{ \int dE \mathcal{F}(E) e^{-iE t} } + \overline{R}\\
		\mathcal{F}(E) &= \sum_{m\neq n, \{j, l\} = \pm} |c_m^j|^2 |c_n^l|^2 \delta\del{E - E_m^j + E_n^l}
	\end{split}
\end{align}
where the equilibrium value of the survival probability, $\overline{R}$ is the inverse participation ratio of the initial state, $\ket{\Psi_\omega}$ in the energy eigenbasis and can be written as
\begin{align}
	\label{eq_R_bar_SC}
	\begin{split}
		\overline{R} &= \mean{\sum_{j = \pm}\sum_n^{\frac{N}{2}} |c_n^j|^4}\\
		&= \sum_{j = \pm}\del{1 + j \sin 2\omega}^2 \sum_{n = 1}^{\frac{N}{2}} \mean{\abs{\Phi_n^j(k)}^4} \\ 
		&\approx \frac{3(1+\sin^2 2\omega)}{N}
	\end{split}
\end{align}
where we used the fact that each of the diagonal blocks in a reduced SC matrix contribute ergodic eigenstates of length $\frac{N}{2}$ following Haar distribution. Thus, energy eigenstate components of a random SC matrix follow the Porter-Thomas distribution~\cite{Bogomolny2018, Manna2025} similar to the GOE. Eq.~\eqref{eq_R_bar_SC} implies that $\ket{\Psi_\omega}$ can spread to the entire energy space or a part of it depending on $\omega$, where the maximal (minimal) spread occurs for $\omega = 0$ $\del{\omega = \pm \frac{\pi}{4}}$.

\begin{figure}[t]
	\centering
	\includegraphics[width=\columnwidth]{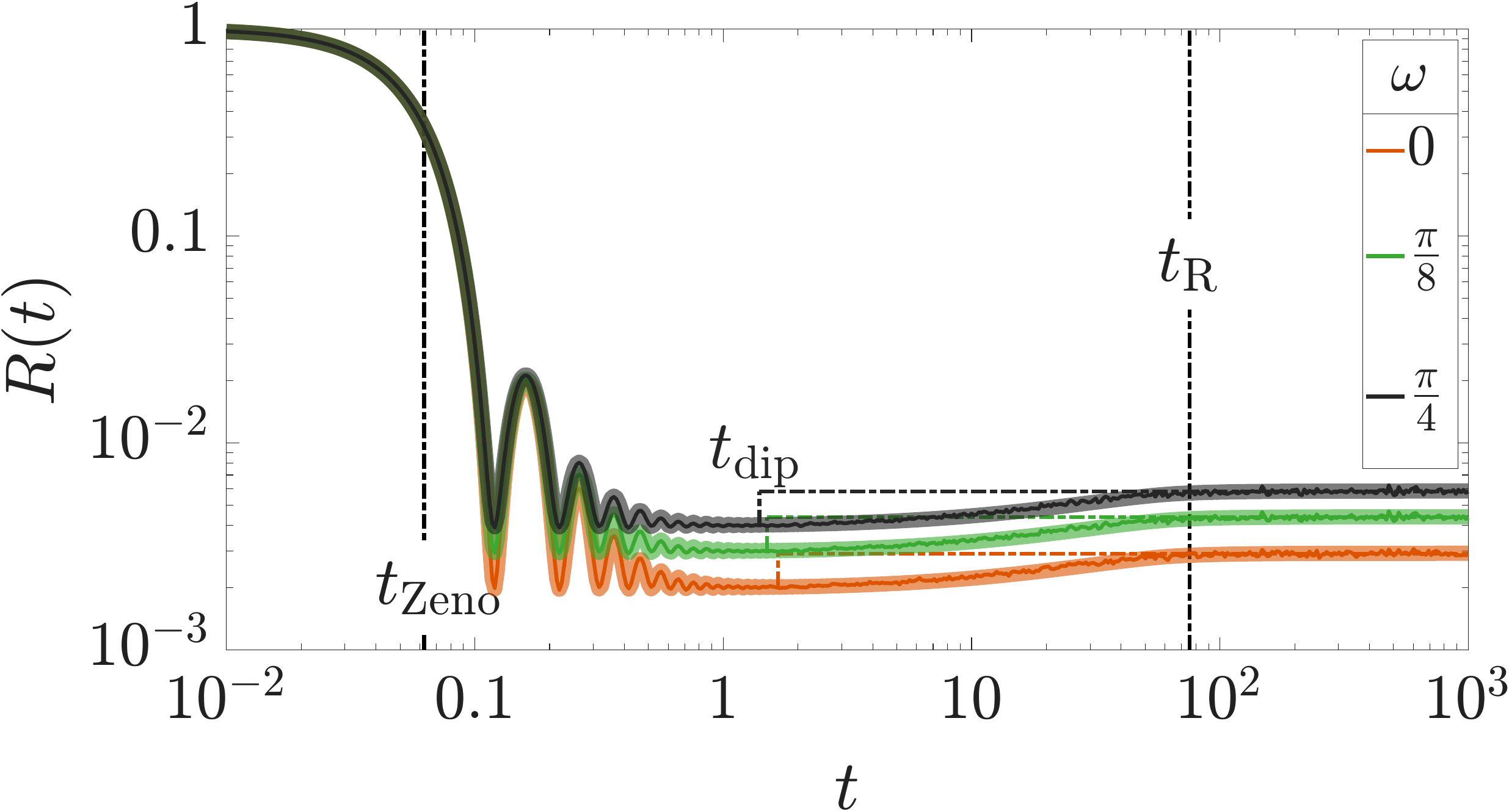}
	\caption[Survival probability for SC matrices]{Ensemble averaged survival probability of $\ket{\Psi_\omega}$ (Eq.~\eqref{eq_ini_state_SC}) for random SC matrices with $N = 1024$ and various values of $\omega$. For survival probability, solid lines denote numerical simulations and the bold curves follow the analytical expression in Eq.~\eqref{eq_R_t_SC}. Vertical black lines denote the Zeno and relaxation times while correlation holes have been marked with dashed lines.}
	\label{fig_SC_SP}
\end{figure}

The 2nd term in Eq.~\eqref{eq_R_t_avg} describes the non-equilibrium evolution of $\ket{\Psi_\omega}$ and can be evaluated following Ref.~\cite{Schiulaz2019} since, the eigenvalues are independent of the eigenstate components in case of a random SC matrix. Then, the survival probability of the initial state in Eq.~\eqref{eq_ini_state_SC} has the following analytical expression
\begin{align}
	\label{eq_R_t_SC}
	\begin{split}
		R_\omega(t) &\approx (1 - \overline{R}) \del{ \frac{\fbslj{1}{2\Gamma t}^2}{(\Gamma t)^2} - \frac{(1+\sin^2 2\omega)}{N} b_2^\mathrm{GOE}\del{\frac{t}{\tH}} }\\ &+ \overline{R}
	\end{split}
\end{align}
where $\fbslj{1}{x}$ is the Bessel function of 1st kind of order 1, $\Gamma$ is the width of the local density of states of $\ket{\Psi_\omega}$, $b_2^\mathrm{GOE}(\tau)$ is given in Eq.~\eqref{eq_b2_GOE}, $\overline{R}$ is given in Eq.~\eqref{eq_R_bar_SC} and $\tau\equiv \frac{t}{\tH}$ is the dimensionless time where $\tH$ is the Heisenberg time
\begin{align}
	\tH \approx \frac{N}{\Gamma} = 2\sqrt{N}.
\end{align}
In Fig.~\ref{fig_SC_SP}, we verify the validity of Eq.~\eqref{eq_R_t_SC} in case of the initial states from Eq.~\eqref{eq_ini_state_SC}. 
For $\omega = \pm \frac{\pi}{4}$, Eq.~\eqref{eq_R_t_SC} describes the survival probability of a localized state in case of a GOE matrix with dimension $\frac{N}{2}$. 

The survival probability exhibits a universal quadratic decay $\sim 1 - \frac{N}{4} t^2$ for $t\ll \tZ$ where the rate of decay is controlled by $\Gamma \approx \frac{\sqrt{N}}{2}$ and $\tZ = \frac{2}{\sqrt{N}}$ is the Zeno time~\cite{Chiu1982}. At later times, $\mean{R_\omega(t)}$ exhibits oscillations with an envelope decaying as $t^{-3}$, a consequence of the energy bounds in the local density of states~\cite{Khalfin1958, Tavora2016, Tavora2017, Roy2025, Cugliandolo2024}. Such a power-law decay (called {\itshape slope}) leads to the dip of the survival probability, which can be found by expanding the 1st and 2nd terms in $R_\omega(t)$ for long and short times, respectively, and finding the minimum of their sum
\begin{align}
	\label{eq_tdip_SC}
	\begin{split}
		& \frac{d}{dt}\del{ \frac{1}{\pi (\Gamma t)^3} - \del{1 - \frac{2\Gamma t}{N}}}\Bigg|_{t = \tdip} = 0\\
		\Rightarrow\: &\tdip = \del{\frac{24}{\pi \del{1 + \sin^2 2\omega}}}^\frac{1}{4}
	\end{split}
\end{align}
Hence, dip time is independent of the system size but depends on the initial state. Beyond $\tdip$, the survival probability exhibits a correlation hole dictated by the two-level form factor, $b_2(\tau)$. At long time, $b_2(\tau)$ has a power-law decay $\del{b_2^\mathrm{GOE}(\tau) \sim \frac{1}{12 \tau^2} }$, thus, we fix a small threshold, $\epsilon$ such that
\begin{align}
	\label{eq_tR_SC}
	\tR = \frac{1}{3}\sqrt{\frac{N}{\epsilon}}.
\end{align}
Hence, relaxation time is independent of $\omega$ and same as that of the GOE~\cite{Schiulaz2019}. Moreover, $\tR$ has the same order of magnitude as that of the Heisenberg time, $\tH \approx 2\sqrt{N}$. In Fig.~\ref{fig_SC_SP}, we show the different timescales for various values of $\gamma$. Eqs.~\eqref{eq_R_bar_SC} and \eqref{eq_tdip_SC} imply that the symmetric $\del{\omega = \frac{\pi}{4}}$ and antisymmetric $\del{\omega = -\frac{\pi}{4}}$ initial states are the quickest ones to reach the correlation hole with minimum spread over the energy space.

\subsection*{Long-lived initial states}
\begin{figure}[t]
	\centering
	\includegraphics[width=\columnwidth]{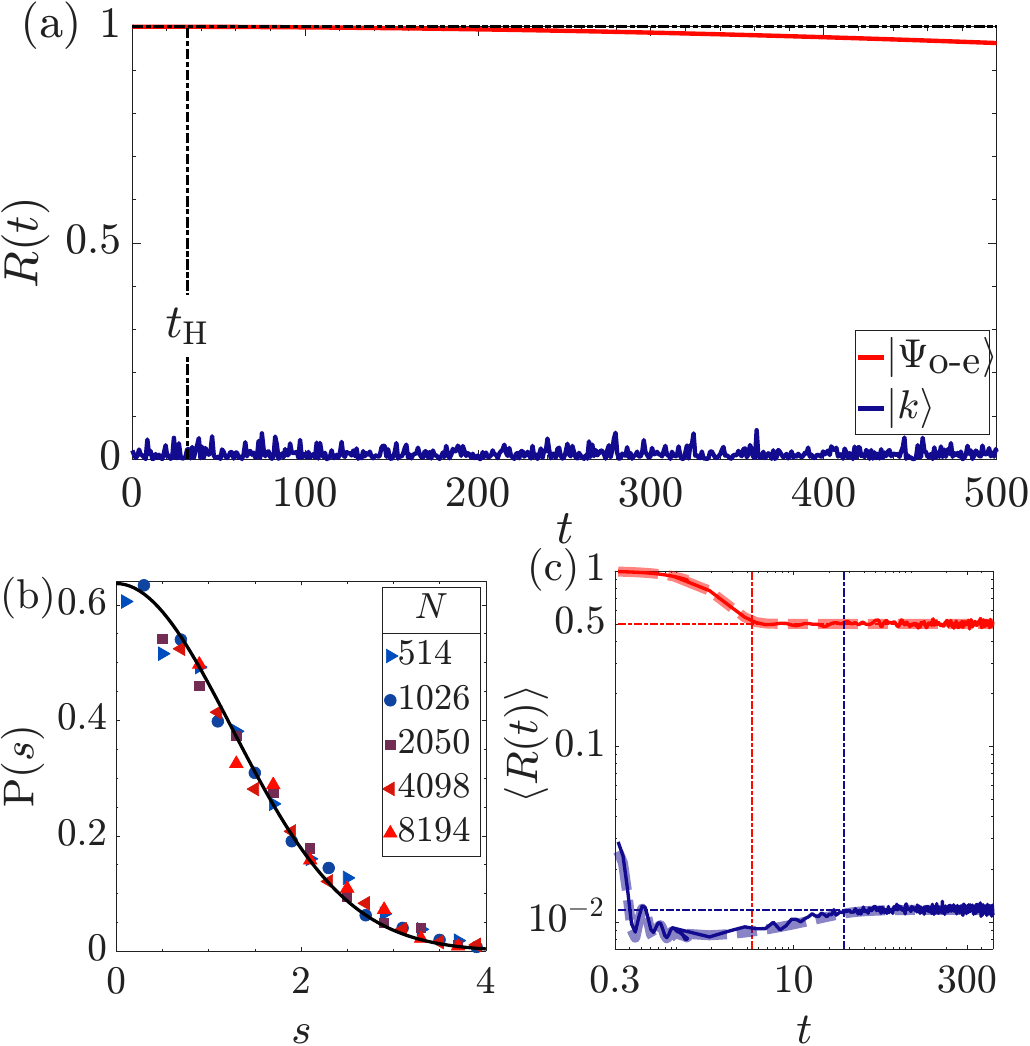}
	\caption[Symmetry-broken initial state for SC matrix]{(a)~Survival probability of the superposition state $\ket{\Psi_\textrm{o-e}}$ (Eq.~\eqref{eq_Rabi_SC}) having energy $\frac{E_1^- + E_1^+}{2}$ and a unit vector $\ket{k}$ with similar energy for a single random SC matrix with $N = 256$. Vertical line denotes the Heisenberg time. Notice that $\ket{\Psi_\textrm{o-e}}$ is almost stationary even up to time $\sim \mathcal{O}(10\tH)$ while $\ket{k}$ equilibrates at $t\sim \tH$.
		(b)~Density of $|E_1^- - E_1^+|$ upon unfolding for different system sizes where the solid line corresponds to the $2\times 2$ Poisson ensemble, Eq.~\eqref{eq:s_o_e}.
		(c)~Ensemble averaged survival probability of $\ket{\Psi_\textrm{o-e}}$ and $\ket{k}$ with respective analytical expressions from Eqs.~\eqref{eq_Rabi_SC_SP_avg} and \eqref{eq_R_t_SC} shown via dashed lines, where the vertical (horizontal) lines denote the relaxation times (equilibrium values).
	}
	\label{fig_SC_SP_sup}
\end{figure}

As the energy spectrum of a random SC matrix is a superposition of two pure spectra from GOE, corresponding ground states, $\ket{\Phi_1^-}$ and $\ket{\Phi_1^+}$ with uncorrelated energies $E_1^-$ and $E_1^+$ (from odd and even sectors, respectively) are the global ground and 1st excited states of the entire spectrum with a high probability, i.e.~lowest two energy levels of a SC matrix are most likely to be $E_1^\pm$. The gap between such energy levels follows the density
\begin{align}
	\prob{s} = \frac{2}{\pi} \exp\del{-\frac{s^2}{\pi}}
	\label{eq:s_o_e}
\end{align}
valid for an integrable two-level system~\cite{Das2019}. If we start from a superposition of the ground states from the two different symmetry sectors, then, the system exhibits Rabi oscillation
\begin{align}
	\label{eq_Rabi_SC}
	\begin{split}
		R(t) &= \abs{ \braket{\Psi_\textrm{o-e}(t)}{\Psi_\textrm{o-e}} }^2 = \cos^2\del{\frac{S_\textrm{o-e}}{2} t}\\
		S_\textrm{o-e} &\equiv |E_1^- - E_1^+|,\quad \ket{\Psi_\textrm{o-e}} \equiv \frac{\ket{\Phi_1^-} + \ket{\Phi_1^+}}{\sqrt{2}}
	\end{split}
\end{align}
with a time period $t_\mathrm{Rabi} = S_\textrm{o-e}^{-1}$. The density of gap between $E_1^-$ and $E_1^+$ (Eq.~\eqref{eq:s_o_e}) implies that they have a high probability to cluster i.e.~$\ket{\Phi_1^-}$ and $\ket{\Phi_1^+}$ can be almost degenerate such that $t_\mathrm{Rabi}$ is arbitrarily large. As a typical excitation takes $\mathcal{O}(\tH)$ time to equilibrate, if 
\begin{align}
	\label{eq_condition_SSB}
	\frac{t_\mathrm{Rabi}}{\tH} \sim \mathcal{O}(N)
\end{align}
then, the time period of the Rabi oscillation is so large that $\ket{\Psi_\textrm{o-e}}$ is effectively an equilibrium state, i.e.~the system retains the memory of the initial configuration for a time period much larger than the Heisenberg time. This is illustrated in Fig.~\ref{fig_SC_SP_sup}(a) for a single realization of a random SC matrix. Thus, similar to the symmetry-broken states in systems exhibiting quantum phase transition, e.g.~transverse field Ising~\cite{Sachdev2011book, Zurek2005}, Lipkin-Meshkov-Glick~\cite{Castanos2006, Wang2021}, collective models~\cite{Hwang2015, Kloc2018, Corps2022}, $\ket{\Psi_\textrm{o-e}}$ breaks the exchange symmetry of a random SC matrix whenever the condition in Eq.~\eqref{eq_condition_SSB} holds, which has a probability of occurring
\begin{align}
	\int_{0}^{\frac{1}{N}} ds\prob{s} = \ferf{\frac{1}{\sqrt{\pi} N}}.
\end{align}
Hence, for a given SC matrix with finite system size, there is a finite probability $\sim \ferf{\frac{1}{\sqrt{\pi} N}}$ that $\ket{\Psi_\textrm{o-e}}$ behaves like a symmetry-broken state, i.e.~an equilibrium state despite not being an energy eigenstate. In the thermodynamic limit ($N\to \infty$), $\ferf{\frac{1}{\sqrt{\pi} N}} = \frac{2}{\pi N} + \mathcal{O}(N^{-3})$ goes to zero such that the probability to find a SC matrix with a symmetry-broken state vanishes. Thus, spontaneous symmetry breaking occurs only for a vanishing fraction ($\sim\mathcal{O}(N^{-1})$) of all possible random SC matrices.

The ensemble averaged survival probability of $\ket{\Psi_\textrm{o-e}}$ can be expressed as
\begin{align}
	\label{eq_Rabi_SC_SP_avg}
	\begin{split}
		\mean{R(t)} &= \int_{0}^{\infty} dS_\textrm{o-e} \prob{S_\textrm{o-e}} R(t)\\
		&= \int_{0}^{\infty} ds \frac{2}{\pi} e^{-\frac{s^2}{\pi}} \cos^2\del{\frac{\mean{S_\textrm{o-e}} s}{2} t}\\
		&= \frac{1}{2} + \frac{1}{2}\exp\del{ -\frac{\pi \mean{S_\textrm{o-e}}^2}{4} t^2 }
	\end{split}
\end{align}
which exhibits a Gaussian decay with a rate $\frac{\sqrt{\pi} \mean{S_\textrm{o-e}}}{2}$ before saturating to $\overline{R} = \frac{1}{2}$. Such Gaussian decay relaxes at the time $\tR  = \frac{2}{\mean{S_\textrm{o-e}}} \sqrt{\frac{|\ln \epsilon|}{\pi}}$ given a tolerance value $\epsilon\ll 1$. Since $\mean{S_\textrm{o-e}}$ is smaller than the global bandwidth $\Gamma$, comparing with Eq.~\eqref{eq_R_t_SC}, we find that the average survival probability of $\ket{\Psi_\textrm{o-e}}$ has a much slower decay but a smaller relaxation time compared to a localized state with similar energy, as shown in Fig.~\ref{fig_SC_SP_sup}(c).

\section[Thermalization]{Equilibration in SC ensemble}\label{sec:SC_thermal}

So far, we looked at the unitary time evolution of an initial state, obtained corresponding survival probability and associated timescales for a random SC matrix. In our analyses, averaging the dynamics over an ensemble ensured equilibration at sufficiently long time due to decoherence~\cite{MatsoukasRoubeas2023}. However, the Poincar\'e recurrence theorem dictates that for a single Hamiltonian of finite dimension, the time-reversal invariance of a unitary dynamics for a discrete spectrum leads to recurrences with a time period proportional to Heisenberg timescale~\cite{Bocchieri1957, Schulman1978, Haake2010book}. 
Such quantum revivals~\cite{Eberly1980} have been demonstrated in the classical limit of the hydrogen atom~\cite{Gaeta1990}. Nevertheless, closed quantum systems can attain equilibrium in the sense that given a local operator, $\ot$, its expectation value can remain arbitrarily close to the infinite time average for a finite interval of time beyond the relaxation time until a recurrence occurs~\cite{Reimann2008}. Particularly, the quantity $|\mean{\ot(t)} - \Tr{\overline{\rho} \ot}|$ can remain arbitrarily small in the said time window where
\begin{align}
	\label{eq_time_avg}
	\overline{\rho} \equiv \lim\limits_{T\to \infty} \frac{1}{T} \int_{0}^{T} dt \ket{\Psi(t)}\bra{\Psi(t)}
\end{align}
is the equilibrium density matrix possessing maximum possible entropy subject to all the conserved charges of the system~\cite{Gemmer2003, Gogolin2011}. 
Such equilibration is rigorously proven for many-body systems provided the local correlations w.r.t.~the initial state decays sufficiently fast~\cite{Cramer2008, Cramer2010}.

To study the relaxation of an observable w.r.t.~a SC matrix, we initialize our system in the state $\ket{\Psi_\omega}$ (Eq.~\eqref{eq_ini_state_SC}) and look at the time evolution of the expectation value of an observable, $\ot$
\begin{align}
	\label{eq_O_t_def}
	\begin{split}
		\ot(t) &\equiv \bra{\Psi_\omega(t)}\ot \ket{\Psi_\omega(t)}\\
		&= \sum_{j, l = \pm} \sum_{m\neq n} (c_m^j)^\star c_n^l \bra{\Phi_m^j} \ot \ket{\Phi_n^l} e^{-i(E_m^j-E_n^l)t} \\&+ \sum_{j = \pm}\sum_n |c_n^j|^2 \bra{\Phi_n^j} \ot \ket{\Phi_n^j}
	\end{split}
\end{align}
where $c_n^\pm \equiv \bra{\Phi_n^\pm}\ket{\Psi_\omega}$. Eq.~\eqref{eq_time_avg} implies that the expectation value of the observable equilibrates to the infinite time average
\begin{align}
	\label{eq_avg_DE_def}
	\mean{\ot}_\mathrm{DE} \equiv \sum_{j = \pm} \sum_n |c_n^j|^2 \ot_{nn}^j, \quad \ot_{nn}^\pm \equiv \bra{\Phi_n^\pm} \ot \ket{\Phi_n^\pm}
\end{align}
which is also known as the average w.r.t.~the diagonal ensemble~\cite{Rigol2007, Cassidy2011, Vidmar2016}. As equilibration is constrained by the conserved quantities, we need to consider all the constants of motion to describe the equilibration process. In a system governed by a time-independent Hamiltonian, the energy is always conserved. For example, $\ket{\Psi_\omega}$, the initial state in Eq.~\eqref{eq_ini_state_SC} has an energy
\begin{align}
	\label{eq_E_ini_state}
	\mean{E} \equiv \bra{\Psi_\omega} H \ket{\Psi_\omega} = \sum_{j = \pm}\sum_n \del{1 + j \sin 2\omega} |\Phi_n^j(k)|^2 E_n^j
\end{align}
which is conserved at all times, during the non-equilibrium evolution as well as beyond $\tR$ when $\ket{\Psi_\omega(t)}$ equilibrates to the maximum entropy state.

In the absence of any conserved quantity other than energy, e.g.~in a completely chaotic system, we expect ergodicity to hold, i.e.~at least the bulk energy states are uniformly distributed on a hypersphere. In a classical system, ergodicity implies the equivalence of the infinite time average and the microcanonical average over the constant energy surface in the phase space~\cite{Bunimovich1979, Simanyi2004, Simanyi2009}. Similarly in an ergodic quantum system, {\itshape typicality} arguments~\cite{Tasaki1998, Popescu2006, Goldstein2006, Linden2009, Mueller2015} imply that the infinite time average of $\ot$ coincides with the average w.r.t.~microcanonical ensemble, which in case of SC matrix is
\begin{align}
	\label{eq_avg_ME_def}
	\mean{\ot}_\mathrm{ME} = \frac{1}{N_\Delta} \sum_{E_n^+, E_n^- \in \Delta} \ot_{nn}^- + \ot_{nn}^+
\end{align}
where $\Delta \equiv [\mean{E}-\Delta_E, \mean{E}+\Delta_E]$ is a microcanonical energy window/ shell around the energy of the initial state containing $1 \ll N_\Delta \ll N$ energy levels over which the density of states varies sufficiently slowly. Importantly, the microcanonical average at a given energy is independent of the choice of the initial state. The equivalence of microcanonical and diagonal averages is shown in quantum systems with few degrees of freedom~\cite{Feingold1985, Feingold1986}.

As the local operator $\ot$ acts on a small part $\mathcal{H}_A$ of the entire Hilbert space, the complimentary subspace acts as a heat bath for $\mathcal{H}_A$~\cite{Lax1955}. Then, the microcanonical average is equal to the canonical expectation value~\cite{DAlessio2016, Mori2018}
\begin{align}
	\label{eq_avg_GE_def}
	\mean{\ot}_\mathrm{GE} \equiv \Tr{\rog \hat{\ot}}, \qquad \rog \equiv \frac{e^{-\frac{\hat{H}}{T}}}{\Trace [e^{-\frac{\hat{H}}{T}}] }
\end{align}
where $\rog$ is the equilibrium thermal density matrix with a canonical partition function $\Trace[e^{-\frac{\hat{H}}{T}}]$ and $T$ is the effective temperature of the initial state following conservation of energy in Eq.~\eqref{eq_E_ini_state}
\begin{align}
	\label{eq_T_eff_ini_state}
	\begin{split}
		&\mean{E} = \Tr{\rog \hat{H}} = \frac{\Tr{ \hat{H} e^{-\frac{\hat{H}}{T}} }}{\Tr{ e^{-\frac{\hat{H}}{T}} }}.
	\end{split}
\end{align}
Thus, Gibbs ensemble allows one to assign a temperature to an initial state~\cite{Santos2010} which determines the size of the microcanonical energy window around $\mean{E}$~\cite{Deutsch2018}. 
Note that in many-body systems with U(1) symmetry, the Gibbs ensemble is defined with an additional Lagrange multiplier for the particle number conservation~\cite{Vidmar2016}. In our case, the SC matrix can be understood as a matrix model of a single-particle closed system where particle number is trivially conserved and becomes the identity matrix in the site basis.

\begin{figure}[t]
	\centering
	\includegraphics[width=\columnwidth]{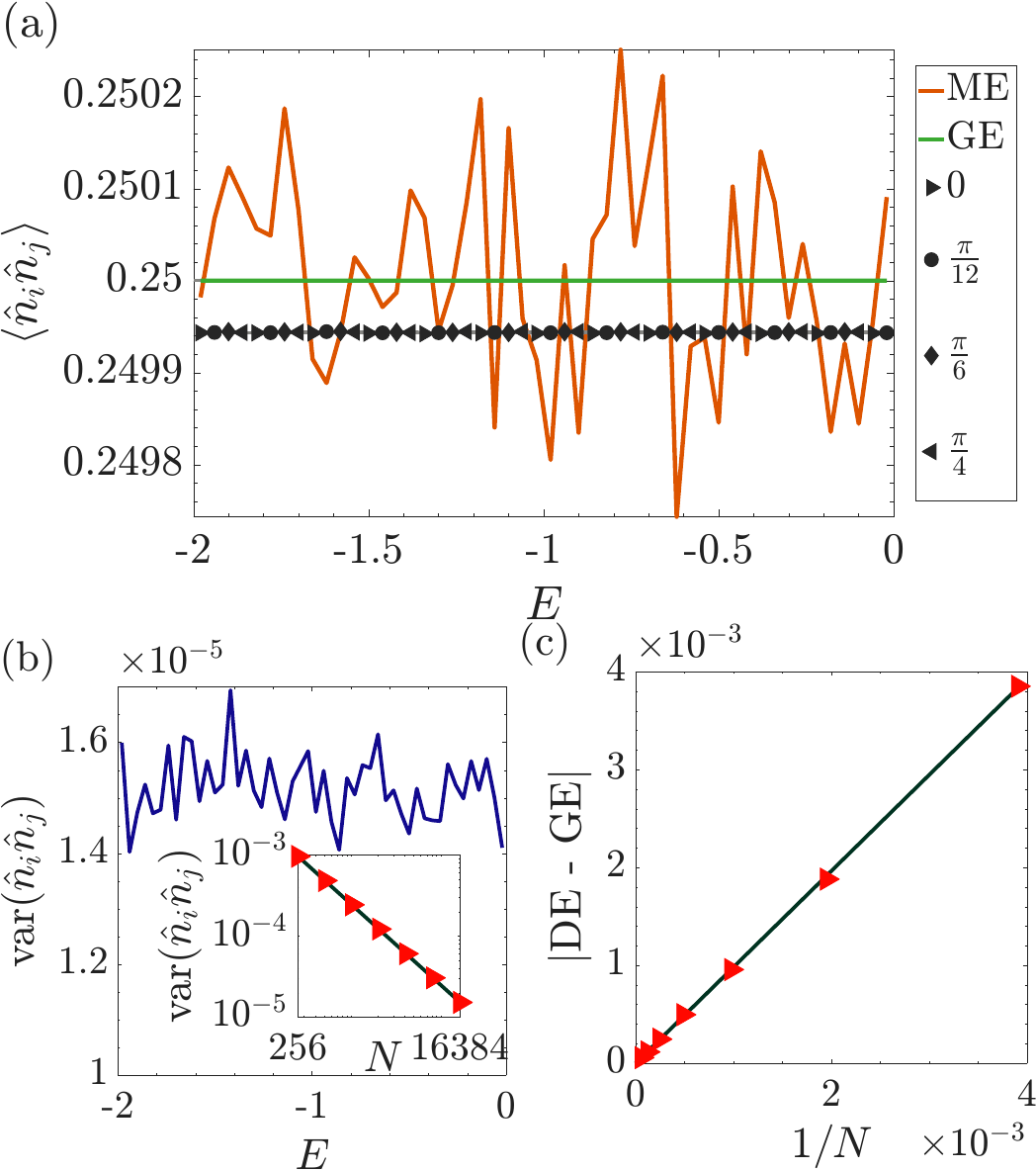}
	\caption{(a)~Expectation of density-density correlation (Eq.~\eqref{eq:ninj_def}) w.r.t.~microcanonical ensemble (ME) and Gibbs ensemble (GE) for $N = 16384$, $i = 1$, $j = \frac{L}{2}$ ($N = 2^L$) averaged over 256 disorder realizations. Microcanonical average is done over windows of width 0.04 in the energy unit. Markers denote the average w.r.t.~diagonal ensemble (DE) for different initial states parameterized by $\omega$ (Eq.~\eqref{eq_ini_state_SC}).
		(b)~Variance of the diagonal terms of the density-density correlation in the energy basis within microcanonical windows. Inset shows the variance w.r.t.~system size where the solid line is linear fit in log-log scale with slope $\sim -1$.
		(c)~Difference of the averages w.r.t.~diagonal and Gibbs ensembles as a function of inverse system size. Solid line denotes linear fit with intercept $\sim 0$.
	}
	\label{fig_nn_avg}
\end{figure}

In case of highly excited initial states and observables not respecting the exchange symmetry, energy conservation alone (Eq.~\eqref{eq_avg_GE_def}) is sufficient to get the thermal expectation value describing the equilibrium behavior. However, in general, the presence of a global symmetry implies that the Gibbs ensemble needs an additional Lagrange multiplier other than the inverse temperature to describe the maximum entropy density matrix~\cite{Rigol2007, Cramer2008, Gogolin2011, Cassidy2011, Caux2013, Mussardo2013, Vidmar2016, Mallayya2021}
\begin{align}
	\label{eq_avg_GGE_def}
	\begin{split}
		\mean{\ot}_\mathrm{GE} &\equiv \Tr{\rog \hat{\ot}},\qquad \rog \equiv \frac{e^{-\frac{H - \mu \J}{T}}}{\nrm}\\
		\nrm &= \Tr{e^{-\frac{H - \mu \J}{T}}} = \sum_{j = \pm}\sum_{n = 1}^{\frac{N}{2}} e^{-\frac{E_n^j-j\mu}{T}}
	\end{split}
\end{align}
where $\nrm$ is the partition function. The Lagrange multipliers $\mu$ and $T^{-1}$ are constrained by the conservation of energy and exchange symmetry, hence, satisfy the following two equations
\begin{align}
	\label{eq_T_eff_GGE}
	\begin{split}
		\mean{E} &= \Tr{\rog \hat{H}} = \frac{ \sum\limits_{j = \pm }\sum\limits_{n = 1}^{\frac{N}{2}} E_n^je^{-\frac{E_n^j-j\mu}{T}}}{\nrm}\\
		\mean{\J} &= \Tr{\rog \hat{\J}} = \frac{ \sum\limits_{j = \pm }j\sum\limits_{n = 1}^{\frac{N}{2}} e^{-\frac{E_n^j-j\mu}{T}} }{ \nrm }
	\end{split}
\end{align}
where $\mean{E}$ is energy (Eq.~\eqref{eq_E_ini_state}) and
$\mean{\J} = \sin 2\omega$ (Eq.~\eqref{eq:J_avg}) is the expectation value of the exchange operator w.r.t~the initial state, $\ket{\Psi_\omega}$. Expanding in the energy eigenbasis, we can express the thermal average $\mean{\ot}_\mathrm{GE}$ as
\begin{align}
	\mean{\ot}_\mathrm{GE} = \frac{ \sum\limits_{j = \pm }\sum\limits_{n = 1}^{\frac{N}{2}} \ot_{nn}^j e^{-\frac{E_n^j-j\mu}{T}} }{ \nrm }.
	\label{eq_avg_GGE_def1}
\end{align}
In integrable and non-ergodic systems, extensively many conserved charges generalize the Gibbs ensemble to have multiple Lagrange multipliers, as demonstrated in many-body systems~\cite{Calabrese2011, Essler2012, Fagotti2013, Fagotti2013a, Fagotti2014, Mierzejewski2014, Pozsgay2014, Ilievski2015, Gramsch2012, Wright2014}, field theories~\cite{Calabrese2007, Fioretto2010, Mussardo2013, Cardy2016}.

The equivalence of the infinite time average from Eq.~\eqref{eq_avg_DE_def} to the canonical average in Eq.~\eqref{eq_avg_GGE_def1} implies {\itshape thermalization}, i.e.~the equilibrium state remains arbitrarily close to the thermal state, $\rog$ beyond relaxation time. This motivated the eigenstate thermalization hypothesis (ETH)~\cite{Srednicki1994, Rigol2008, Polkovnikov2011a, Nandkishore2015, DAlessio2016, Borgonovi2016, Mori2018, Deutsch2018}, which dictates that in an ergodic closed quantum system with eigenpairs $\cbr{E_n, \Phi_n}$, the matrix elements of an observable follow the ansatz
\begin{align}
	\bra{\Phi_m} \hat{\ot} \ket{\Phi_n} = \ot(\overline{E})\delta_{m n} + \frac{1}{\sqrt{\rho(\overline{E})}} \mathcal{F}(\overline{E}, \omega) \mathcal{R}_{mn}
	\label{eq:ETH_ansatz}
\end{align}
where $\overline{E} = \frac{E_m + E_n}{2}$, $\omega = E_n - E_m$, $\ot(\overline{E})$ and $\mathcal{F}(\overline{E}, \omega)$ are smooth functions of their arguments, $\rho(E)$ is the density of states and $\mathcal{R}_{mn}$ is a random number with zero mean and unit variance. The ETH ansatz also implies that the microcanonical (Eq.~\eqref{eq_avg_ME_def}), canonical (Eq.~\eqref{eq_avg_GGE_def}) and equilibrium (Eq.~\eqref{eq_avg_DE_def}) expectation values of a generic local observable coincide. Note that, despite being integrable, certain translationally invariant finite range quantum systems may satisfy ETH in the weak sense~\cite{Mueller2015}.

To probe the validity of ETH in case of a random SC matrix, we obtain the expectation values of various local observables after initializing our system in a state $\ket{\Psi_\omega}$ (Eq.~\eqref{eq_ini_state_SC}). The density of states of a SC matrix is symmetric about the zero energy, hence, the energy of the initial state (Eq.~\eqref{eq_E_ini_state}) is $\sim\mathcal{O}(1)$, lies in the middle of the spectrum while the spectral bandwidth scales as $\mathcal{O}(\sqrt{N})$. Thus, our chosen initial states are highly excited. The semicircle law implies that within the energy window $\mean{E} \pm \Delta_E$ around zero energy, there exists $\mathcal{O}(\Delta_E \sqrt{N})$ number of energy levels. Thus, such an energy window can be considered to be a microcanonical shell. In our numerical results, we choose $\Delta_E = 0.02$.
	
First, we look at the correlation between the number densities at the $i$th and $j$th sites of a fictitious 1D lattice
\begin{align}
	\hat{\ot} = \hat{n}_i \hat{n}_{j},\quad \hat{n} = \frac{\hat{\mathbb{I}} - \hat{\sigma^z}}{2}
	\label{eq:ninj_def}
\end{align}
assuming that the matrix $H$ is the Fock space representation of a spin-$\frac{1}{2}$ chain with $L$ number of sites, $N = 2^L$. Such a local operator does not commute with the exchange matrix $\J$. As the eigenstates are Haar random vectors other than the $\mathbb{Z}_2$ symmetry, we expect that the site occupations are homogeneous over the lattice, $\mean{\hat{n}_i} \approx \frac{1}{2}$ for all lattice indices and independent of each other. Then, $\mean{\hat{n}_i \hat{n}_{j}} = \mean{\hat{n}_i} \mean{\hat{n}_j} \approx \frac{1}{4}$ irrespective of the choice of initial states.

\begin{figure}[t]
	\centering
	\includegraphics[width=\columnwidth]{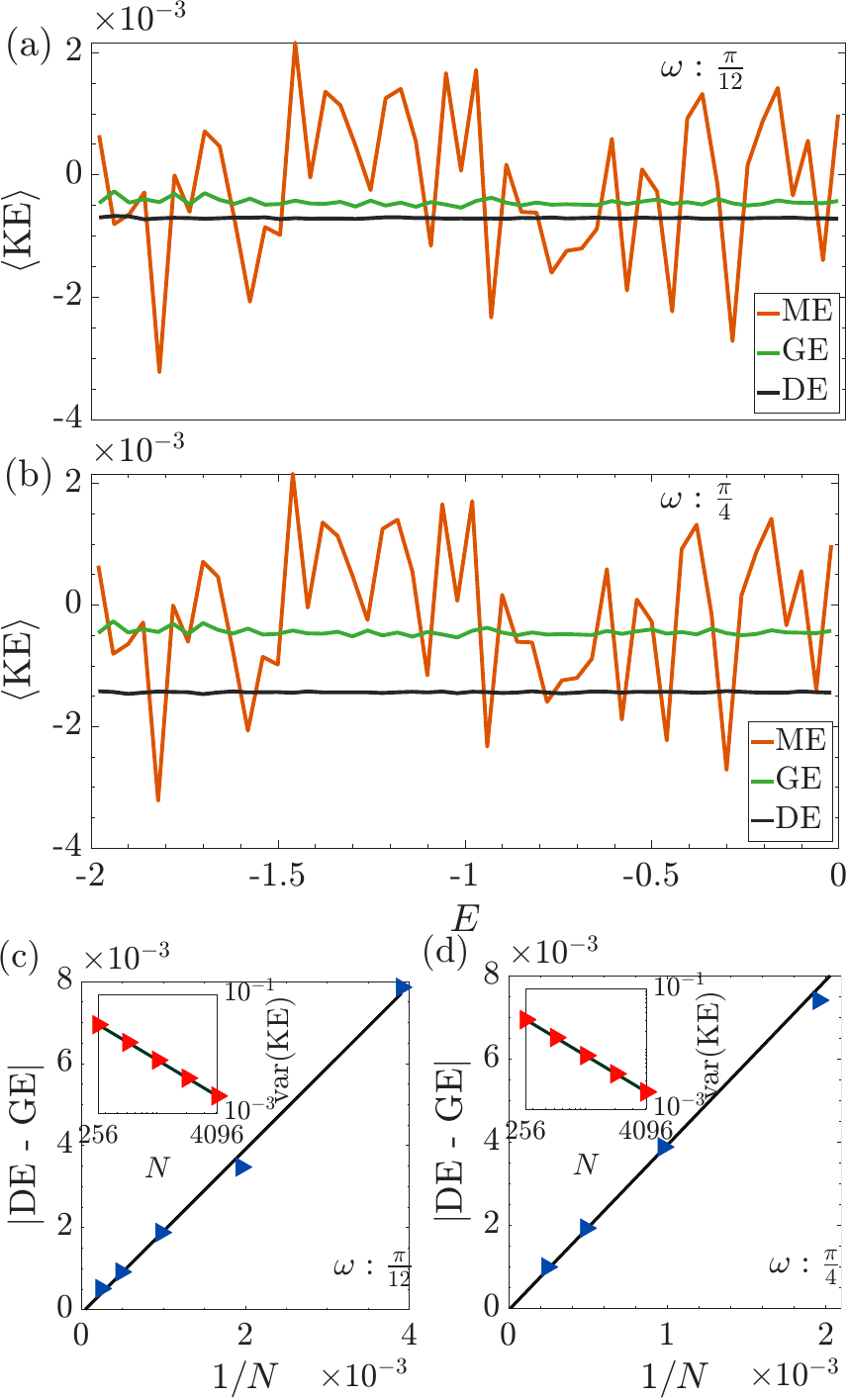}
	\caption{Expectation of kinetic energy operator (Eq.~\eqref{eq:KE_def}) w.r.t.~microcanonical ensemble (ME), Gibbs ensemble (GE) and digonal ensemble (DE) for (a)~$\omega = \frac{\pi}{12}$ and (b)~$\omega = \frac{\pi}{4}$ with $N = 4096$ averaged over 1024 disorder realizations. $\omega$ characterizes the initial state (Eq.~\eqref{eq_ini_state_SC}).
		Difference of the averages w.r.t.~diagonal and Gibbs ensembles as a function of inverse system size for (c)~$\omega = \frac{\pi}{12}$ and (d)~$\omega = \frac{\pi}{4}$. Solid line denotes linear fit with intercept $\sim 0$. Insets show the variance of the diagonal terms of the kinetic energy operator in the energy basis within microcanonical windows where the solid line is linear fit in log-log scale with slope $\sim -1$.
	}
	\label{fig_KE_avg}
\end{figure}

In Fig.~\ref{fig_nn_avg}(a), we show the expectation value of the density-density correlation w.r.t.~different ensembles as a function of the energy of $\ket{\Psi_\omega}$. As the observable does not respect the global symmetry of the Hamiltonian, the equilibrium value (average w.r.t.~diagonal ensemble) does not depend on the choice of the initial state, as shown for different values of $\omega$ in Fig.~\ref{fig_nn_avg}(a). The microcanonical average fluctuates around the expected value of $\mean{\hat{n}_i \hat{n}_{j}} = \frac{1}{4}$. In Fig.~\ref{fig_nn_avg}(b), we show that the variance of the diagonal terms of the density-density correlation within microcanonical shells decays linearly with the system size. Thus, diagonal ETH ansatz (Eq.~\eqref{eq:ETH_ansatz}) is valid for our observable.

As the energy of the initial state is close to the middle of the spectrum, we expect equipartition of energy states with Gibbs weight $\sim N^{-1}$ such that the thermal average must coincide with the microcanonical average, as shown in Fig.~\ref{fig_nn_avg}(a). We find a small difference between the averages w.r.t.~diagonal and canonical ensembles for finite system sizes. Such a difference decays linearly with $N^{-1}$ and vanishes in the limit $N\to \infty$, as shown in Fig.~\ref{fig_nn_avg}(c). Note that as $\hat{n}_i \hat{n}_{j}$ do not commute with $\J$, energy conservation alone is sufficient to get the thermal expectation values as in Eq.~\eqref{eq_avg_GE_def}. Thus, we find that the expectation values w.r.t.~microcanonical, canonical and diagonal ensembles coincide for the density-density correlation, validating thermalization.

Next, we look at the kinetic energy operator of 1D single-particle systems~\cite{Lydzba2021}
\begin{align}
	\hat{\mathrm{KE}} = - \sum_{k = 1}^{N-1} \hat{c}_k^\dagger \hat{c}_{k+1} + \hat{c}_{k+1}^\dagger \hat{c}_k
	\label{eq:KE_def}
\end{align}
where $\hat{c}_j^\dagger$ ($\hat{c}_j$) is the creation (annihilation) operator on $j$th site of a 1D tight-binding model. In the site basis, $\cbr{\ket{k}}$, the above operator is tridiagonal with zero diagonal. Importantly, $\hat{\mathrm{KE}} \ket{k} = \ket{N - k + 1}$ such that $[\hat{\mathrm{KE}}, \J] = 0$.

In Fig.~\ref{fig_KE_avg}(a) and (b), we show the expectation value of the kinetic energy w.r.t.~various ensembles for two different values of $\omega$. The average w.r.t.~Gibbs ensemble is computed following the conservation of both energy and exchange symmetry as in Eq.~\eqref{eq_avg_GGE_def}. As the observable respects the global symmetry of the Hamiltonian, the equilibrium value depends on the choice of the initial state, unlike the density-density correlation. While for finite system sizes, we see a difference between the canonical average and equilibrium value, such a difference decays linearly with system size as shown in Fig.~\ref{fig_KE_avg}(c) and (d). Moreover, the fluctuations of the diagonal terms within microcanonical windows also vanish in the thermodynamic limit on par with the ETH ansatz, as shown in the insets of Fig.~\ref{fig_KE_avg}(c) and (d).

\begin{figure}[t]
	\centering
	\includegraphics[width=\columnwidth]{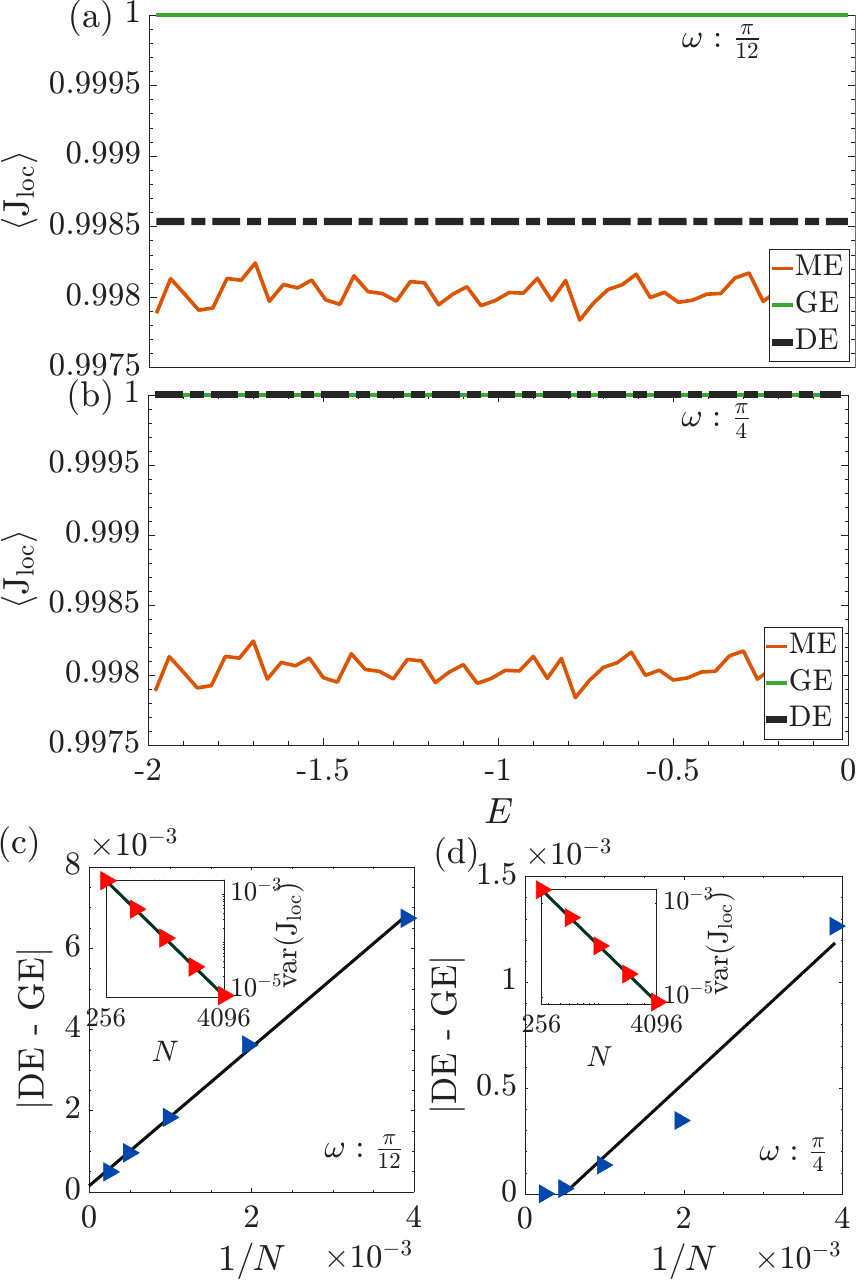}
	\caption{Expectation of the operator in Eq.~\eqref{eq:Jloc_def} w.r.t.~microcanonical ensemble (ME), Gibbs ensemble (GE) and digonal ensemble (DE) for (a)~$\omega = \frac{\pi}{12}$ and (b)~$\omega = \frac{\pi}{4}$ with $N = 4096$ averaged over 1024 disorder realizations. $\omega$ characterizes the initial state (Eq.~\eqref{eq_ini_state_SC}).
		Difference of the averages w.r.t.~diagonal and Gibbs ensembles as a function of inverse system size for (c)~$\omega = \frac{\pi}{12}$ and (d)~$\omega = \frac{\pi}{4}$. Solid line denotes linear fit with intercept $\sim 0$. Insets show the variance of the diagonal terms of the kinetic energy operator in the energy basis within microcanonical windows where the solid line is linear fit in log-log scale with slope $\sim -1$.
	}
	\label{fig_Jloc_avg}
\end{figure}

Thus, kinetic energy operator also exhibits thermalization in case of random SC matrices despite respecting the global symmetry of the Hamiltonian. Similar behavior is observed in case of the following operator
\begin{align}
	\J_\mathrm{loc} = \mathbb{I}^{\frac{N-4}{2}\times \frac{N-4}{2}} \oplus \J^{4\times 4} \oplus \mathbb{I}^{\frac{N-4}{2}\times \frac{N-4}{2}}.
	\label{eq:Jloc_def}
\end{align}
which commutes with the global exchange symmetry and exhibits thermalization, as shown in Fig.~\ref{fig_Jloc_avg}.

\section{Discussion}\label{sec:discuss}

In this work, we look at the random SC matrix which has a discrete $\mathbb{Z}_2$ symmetry and commutes with the exchange symmetry operator. Corresponding Hilbert space can be split into two decoupled subspaces, namely odd (spanned by antisymmetric states) and even (spanned by symmetric states) sector. The energy spectrum of a SC matrix is a superposition of two pure spectrum, hence, we can analytically describe the energy correlations, e.g.~the level spacing distribution and number variance.

To understand the dynamical properties of random SC matrices, we consider the initial state $\ket{\Psi_\omega}$, which is a function of $-\frac{\pi}{2} \leq \omega \leq \frac{\pi}{2}$ and a weighted superposition of the exchange doublets. Depending on the value of $\omega$, the initial state can be confined in the odd or even sector or can couple these subspaces as well. We analytically obtain the survival probability (i.e.~the probability of the system returning to its initial configuration) of $\ket{\Psi_\omega}$, corresponding equilibrium value and characteristic timescales. The survival probability exhibits a correlation hole for all values of $\omega$ indicating the presence of long-range correlation in the energy spectrum. However, the symmetric ($\omega = \frac{\pi}{4}$) and antisymmetric ($\omega = -\frac{\pi}{4}$) states take the smallest amount of time to reach the correlation hole and have the minimal spread over the energy space. Next, we look at the initial state $\ket{\Psi_\textrm{o-e}}$, which is a superposition of the local ground states of the two decoupled subspaces. We show that $\ket{\Psi_\textrm{o-e}}$ is a symmetry-broken state for a measure zero fraction of random SC matrices with spontaneous symmetry breaking in the thermodynamic limit.

To understand the thermalization in case of SC matrices, we consider initial states and local observables which do and do not respect the global symmetry of the Hamiltonian. By comparing the equilibrium expectation value, microcanonical and canonical averages of the local observables, we show that thermalization takes place in case of random SC matrices. We show that the equilibrium behavior is dependent on the choice of initial state if the observable respects the global symmetry of the Hamiltonian and well described by the Gibbs ensemble for all the choices of initial state and observables. 
	
As a future direction, we would like to explore \textit{deep thermalization}~\cite{Ippoliti2022, Sherry2026Arxiv} in case of SC matrices, which is a stronger notion of thermalization where the projections of a time evolved initial state spread uniformly over a subspace. It has been shown that in the presence of a global symmetry (e.g.~Sachdev-Ye-Kitaev model), higher moments of the projected ensemble can deviate from the Haar ensemble, indicating the absence of deep thermalization despite ETH remaining valid~\cite{Bhore2023}. The SC matrix can pose as a minimal toy model to exhibit this phenomena. One can also look at the dynamics of $\ket{\Psi_\omega}$ and thermalization in case of the deformed centrosymmetric ensemble~\cite{Das2022b}, where the extent of centrosymmetry can be tuned. The dependence of the relaxation times of various initial states on the strength of symmetry breaking perturbation can provide important insight in the equilibration of disordered systems.

\begin{acknowledgements}
	We sincerely thank Anandamohan Ghosh and Achilleas Lazarides for useful discussions and constructive criticism of the manuscript. A.K.D.~acknowledges the support from the Leverhulme Trust Research Project Grant RPG-2025-063. All the computations were performed in the Kepler Computing facility of the Department of Physical Sciences, IISER Kolkata, India. All the data shown in this work are publicly available in~\footnote{\url{https://github.com/AdwayDas/Symmetric\_centrosymmetric\_data}}.
\end{acknowledgements}

\bibliography{ref_SC}

\end{document}